\documentclass{article}
\usepackage{lnfprep}
\usepackage{graphicx}
\usepackage{subfigure}
\usepackage{amsmath}
\usepackage{amssymb}
\def\beq{\begin{equation}}   \def\eeq{\end{equation}}
\def\bea{\begin{eqnarray}}   \def\eea{\end{eqnarray}}

\newcommand{\gsim}{\lower.7ex\hbox{$ \;\stackrel{\textstyle>}{\sim}\;$}}
\newcommand{\lsim}{\lower.7ex\hbox{$ \;\stackrel{\textstyle<}{\sim}\;$}}
\def\c2{CLEO~II.V}

\def\d0d0{ D^0\bar{D}^0 }
\def\p0p0{ P^0\bar{P}^0 }

\def\qp2{ \Bigl| \frac{q}{p} \Bigr|^2 }
\def\pq2{ \Bigl| \frac{p}{q} \Bigr|^2 }

\def\ps2s{  \psi(2S) }
\def\q2{ $q^2$ }
\def\cm2s1{ $\,{\rm cm}^{-2} {\rm s}^{-1}$}
\def\d0{D_2^{*0}}
\def\d+{D_2^{*+}}

\newcommand{\Header}{
  \begin{tabular}{rl}
  \hspace{-.4cm}
      &
    \renewcommand{\arraystretch}{0.5}
    \renewcommand{\arraystretch}{1}
  \end{tabular}
  \vskip 1cm
  \begin{flushright}
  \renewcommand{\arraystretch}{0.5}
    \begin{tabular}{r}
      {\underline{LNF-04/25 (P)}}\\    % insert here the preprint number
      {\small 27 dicembre 2006} \\      % insert here the preprint Date
    \end{tabular}
  \end{flushright}
  \renewcommand{\arraystretch}{1}
  \vskip 1 cm
  }
\begin{document}
\begin{titlepage}
\title{
  \Header
  {\LARGE  \textsc{\textmd {Gas Analysis and Monitoring Systems
for the RPC Detector of CMS at LHC
}}
  }
}
\author{M.Abbrescia$^a$, A.Colaleo$^a$ , R.Guida$^a$, G.Iaselli$^a$, R.Liuzzi$^a$, F.Loddo$^a$, M.Maggi$^a$,\\
B.Marangelli$^a$, S.Natali$^a$, S.Nuzzo$^a$, G.Pugliese$^a$, A.Ranieri$^a$, F.Romano$^a$, R.Trentadue$^a$,\\
L.Benussi$^b$, M.Bertani$^b$, M.A.Caponero$^c$, D.Colonna$^d$, D.Donisi$^c$, F.L.Fabbri$^b$, F.Felli$^d$, M.Giardoni$^b$,\\
B.Ortenzi$^b$, M.Pallotta$^b$, A.Paolozzi$^e$,  L.Passamonti$^b$, B.Ponzio$^b$, C.Pucci$^e$, G.Saviano$^d$,\\
G.Polese$^{f,g}$, I.Segoni$^g$, N.Cavallo$^h$, F.Fabozzi$^h$, P.Paolucci$^i$, D.Piccolo$^i$, C.Sciacca$^i$,\\
G.Belli$^l$, A.Grelli$^l$, M.Necchi$^l$, S.P.Ratti$^l$, C.Riccardi$^l$, P.Torre$^l$, P.Vitulo$^l$}
%{\em Stefano Bianco, on behalf of the CMS RPC Collaboration1 (Bari - %Frascati - Napoli - Pavia)}}
\maketitle
\baselineskip=1pt

\begin{abstract}
\indent The Resistive Plate Chambers (RPC) detector of the CMS experiment at the LHC proton collider (CERN, Switzerland) will employ an online gas analysis and monitoring system of the freon-based gas mixture used. We give an overview of the CMS RPC gas system, describe the project parameters and first results on gas-chromatograph analysis. Finally, we report on preliminary results for a set of monitor RPC.
\end{abstract}

\vspace*{\stretch{2}}
%\begin{flushleft}
% insert here the PACS number
%  \vskip 2cm
%{ PACS.: wire chambers, straw tubes, HEP detectors, silicon
%detectors, silicon microstrips, beauty quark, CP violation.}
%\end{flushleft}
%\begin{center}
%\emph{Submitted to Transactions on Nuclear Science}
%\end{center}

\vskip 1cm
\begin{flushleft}
Manuscript received on November 17, 2006.\\
Stefano Bianco is with the Laboratori Nazionali di Frascati dell'INFN, v.E.Fermi 40, 00044 Frascati (Rome) Italy (telephone: +39-06-94032793, e-mail: stefano.bianco@lnf.infn.it).\\
\vskip 1cm
\begin{tabular}{l l}
  \hline\\
  $ ^{a\,\,\,\,\,\,}$ & \footnotesize{Dipartimento Interateneo di Fisica and Sezione INFN, Bari, Italy;} \\
  $ ^{b\,\,\,}$& \footnotesize{Laboratori Nazionali di Frascati dell'INFN, Italy;} \\
  $ ^{c}$ & \footnotesize{Laboratori Nazionali di Frascati dell'INFN and ENEA Frascati, Italy;} \\
  $ ^{d}$ & \footnotesize{Laboratori Nazionali di Frascati dell'INFN and Facolta' di Ingegneria Roma1, Italy} \\
  $ ^{e}$ & \footnotesize{Laboratori Nazionali di Frascati dell'INFN and Scuola di Ingegneria aerospaziale Roma 1, Italy} \\
  $ ^{f}$ & \footnotesize{CERN, Geneva, Switzerland;} \\
  $ ^{g}$ & \footnotesize{Lappeenranta University of Technology, Finland;} \\
  $ ^{h}$ & \footnotesize{Universita' degli Studi della Basilicata and Sezione INFN, Napoli, Italy;} \\
  $ ^{i}$ & \footnotesize{Dipartimento di Fisica and Sezione INFN, Napoli, Italy;} \\
  $ ^{l}$ & \footnotesize{Dipartimento di Fisica Nucleare e Teorica and Sezione INFN, Pavia, Italy} \\
\end{tabular}
\end{flushleft}
\end{titlepage}
\pagestyle{plain}
\setcounter{page}2
\baselineskip=17pt

\section{\textsc{Introduction}}
Resistive Plate Chambers (RPC) detectors are widely used in HEP experiments for muon detection and triggering at high-energy, high-luminosity hadron colliders, in astroparticle physics experiments for the detection of extended air showers, as well as in medical and imaging applications. While gain and efficiency stability are always a must, in the case of RPC detectors in high-rate experiments which use freon-based gas mixtures, utmost care has to be paid also for the possible presence of gas contaminants.
The RPC detector of experiment CMS at the LHC proton collider (CERN, Switzerland) will employ a gas analysis and monitoring system for the online monitor of the freon-based gas mixture used. The gas monitoring system is based on small RPC detectors whose working point (gain and efficiency) is continuously monitored online. The gas monitoring system is designed to provide fast and accurate determination of any shift in working point conditions. 	Quantitative gas chemical analysis is then performed online by a complete system which includes gas-chromatography, pH sensors and contaminants (notably HF) detectors.
\section{\textsc{The CMS Experiment at LHC}}
The Compact Muon Solenoid (CMS) experiment [1] will search for the missing block of Nature - the Higgs boson - and for new exotic elementary particles that are predicted by theory and by cosmological observations. The CMS detector  uses Resistive Plate Chambers (RPC)  as muon detectors, coupled to Drift Tubes in the barrel region, and to Cathode Strip Chambers in the endcaps. The RPC detector in CMS was described elsewhere at this conference [2].

RPC counters [3] are fast, efficient and economical charged particles detectors, well-suited for operation in high magnetic field. The elementary component is a gap, a gas volume enclosed between two resistive plates. Resistive plates are made of bakelite, coated with lineseed oil for surface uniformity. Gas mix used is 96.2\% C$_2$H$_2$F$_4$ / 3.5\% Iso-C$_4$H$_{10}$ / 0.3\% SF$_6$, with  a 45\% relative humidity.

Signal pulses are picked up by readout strips. In CMS, RPC counters are operated in avalanche mode to sustain high-rate operation, with the streamer suppressed by the addition of SF$_6$ gas in the mixture.

\section{  \textsc{The CMS Closed Loop Gas System}}
Because of high costs and huge volumes of the freon-based gas mix used, CMS will use a recirculation (Closed Loop) gas system developed by the CERN gas group [4]. The Closed Loop is a critical component of RPC. CMS has accumulated experience on its use and performances during the test at the Gamma Irradiation Facility at CERN  in 2001 [5], and currently at the ISR where chambers are tested in CL prior to installation. At the GIF facility we observed substancial production of HF, linearly correlated with the signal current.

In the Closed Loop (CL) system, purifiers are the crucial componente. Purifiers were determined after tests at the GIF in order to minimize the unknown contaminants which showed as spurious peaks besides the known gas mix components. Three filters were selected: 5A molecular sieve, Cu/Cu-Zn, Ni/Al$_2$O$_3$.
A small scale CL system is currently in use at the ISR test station, where RPC chambers are tested at CERN prior to installation in the CMS detector. The system total flux is 110 l/h, with a 10\% fraction of fresh mix. Our operational experience showed how the CL system works well as long as the purifiers are not saturated (about 20 days). When purifiers are saturated, contaminants are not filtered and currents in chambers start increasing (Fig.1). Currents return to standard after purifiers regeneration. We do not observe any trace in µGC analysis of either impurities or pollutants.

A measurement campaign[6] on purifiers is in progress, using chemical, SEM/EDS (Scanning Electron Microscopy/Energy Dispersive Spectroscopy), XRD (x-ray Diffrattometry)  analyses. We plan to characterize the CL gas system in three phases: at ISR during chamber testing, at ISR at testing finished with dedicated gaps and dedicated gas system, at the GIF in high-radiation environment. Many open questions do exist: why only some 30-40\% of chambers show current increase with saturated purifiers? Why often only one gap out of two is affected? It is clear how full understanding of the CL system is crucial for a reliable operation of CMS RPCs.
\newline

\section{  \textsc{Gas Quality Monitoring System}}
Gas quality in CMS RPC will be monitored by a dedicated system able to accomplish a full analysis of the gas quality. The gas quality monitoring system will use specific electrodes for hydro-fluoridric acid (HF) detection, as well as SEM/EDS, XRD analyses for purifiers. The system will be flexible and open, so as to integrate other analysis devices such as GC (Gas Chromatography) and MS (Mass Spectrometry).
In case of anomalies detected, we shall have the possibilities of accessing the experimental cavern where manual pickup points (one per half wheel) will be setup.
Several results have been produced over the last few years with subsystems planned to be used in the gas quality monitoring system. Production of  HF was found in chambers irradiated at the GIF[5], with the concentration of HF produced in the gap found proportional to the signal charge.
We recently performed new studies on irradiated RPCs. We opened  small (50cm x 50cm) chambers irradiated at the GIF and observed defects on the inner surfaces (Fig.2). We performed SEM-EDS analyses on- and off-defect. The presence of Na in defects is confirmed. The origin of Na is supposedly the bakelite bulk, where NaOH is used as a catalyst in production.
We performed XRD analysis of defects, preliminary results [7] show a good match of diffrattogram with the lines characteristic of NaF (Fig.3).
\newline

\section{  \textsc{Gas Gain Monitoring System}}
The gas gain monitoring system will monitor the RPC working point faster and more precisely than what one could get by using the CMS RPC system, and provide a warning in case of shifts caused by the gas mixture changes. The system is designed to monitor efficiency and charge continuously in one-hour cycles with a 1% precision.
The system (Fig.4)  is composed of three subsystem of RPC single gaps, readout by 45cm x 45cm pads in a cosmic ray telescope located in the SGX5 gas building.  Each subsystem is flushed with a different gas. The Reference subsystem is flushed with fresh open loop gas mixture. The MonitorOut subsystem is flushed with CL gas downstream of CMS RPCs. The MonitorIn subsystem is flushed with CL gas upstream of CMS RPCs. Each subsystem is composed of three gaps, whose high voltage is set to the standard working point voltage at the efficiency knee, and to 200V above and below the knee respectively.
Each cosmic ray track therefore provides completely correlated pulses in the three subsystems, allowing one to study the differential response of gaps and by disentangling any effect due to changes in the gas mixture. In case a working point change is detected, an alarm condition is released and the gas quality monitoring system described in Sec.IV will verify what the change of work point is due to.
\subsection{\emph{Prototypes}}
Several readout options have been considered for the RPC single gaps composing each subsystem. A double-pad readout was investigated, with each single pad read by both sides by 45cm x 45cm pads. An exploded view is shown in Fig.5.
\subsection{\emph{Preliminary Results}}
Single gap prototypes were exposed to cosmic ray tracks triggered by a scintillator counter hodoscope. The negative and positive pads are sent to two input channels of a Tektronix TDS5600 digital sampling oscilloscope. Event-by-event pulse charges are measured and stored via LabView custom applications.  The standard CMS RPC gas mixture is used, with 45\% relative humidity to keep the bakelite resistivity.

Fig.6 shows the charge distributions of avalanches from cosmic rays for positive and negative  pads, at high voltages from 9.5kV to 10kV.  Fitting the charge distributions with truncated Gaussians shows the expected linear dependance of charge at the peak of distribution on the high voltage applied in case of saturated avalanche (Fig.7).

Several amplification schemes are being explored. Even a simple passive sum of positive and negative pads, and feeding to NIM LeCroy LRS612AM amplifier improves the S/N ratio (Fig.8).
\newline
\section{  \textsc{Conclusions}}
The CMS RPC group is performing a detailed and complete analysis campaign since early tests at the GIF in 2001, to guarantee high-purity gas mixture for a reliable operation of the detector. A lot of work is being spent into the full understanding of the chemistry of purifiers in CL gas system.
A gas analysis system has been designed, SEM-EDS analysis observes presence of Na in RPC gaps irradiated at GIF confirming previous results, while XRD analysis shows the presence of NaF.
A gas gain monitoring system utilizing small RPC gaps has been designed, prototypes have been tested and preliminary results show the expected response to cosmic rays.
\newline
\newpage

\newpage
\begin{figure}[!htbp]
\begin{center}
  \includegraphics[]{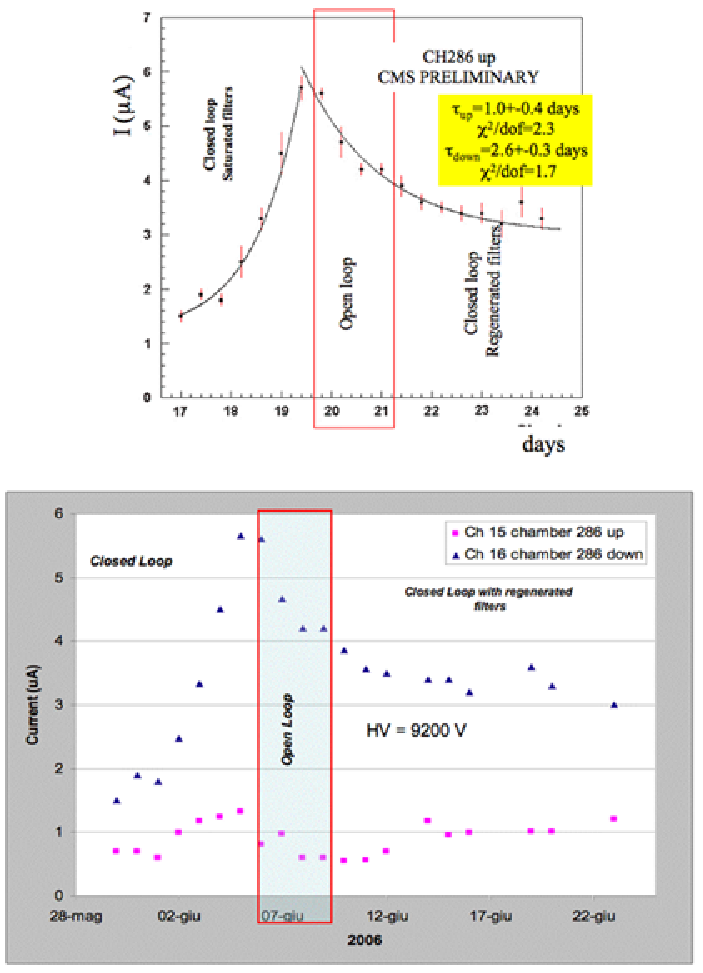}\\
  \caption{Increase of currents in RPC chambers under test at the ISR in CL gas system when purifiers are saturated. Currents start decreasing when chamber is set in Open Loop, and decrease further when chamber is returned to CL with regenerated purifiers. Top inset show results of a best fit of an exponenential curve to data.}
  \end{center}
\end{figure}

\

\begin{figure}[!htbp]
\begin{center}
  \includegraphics[]{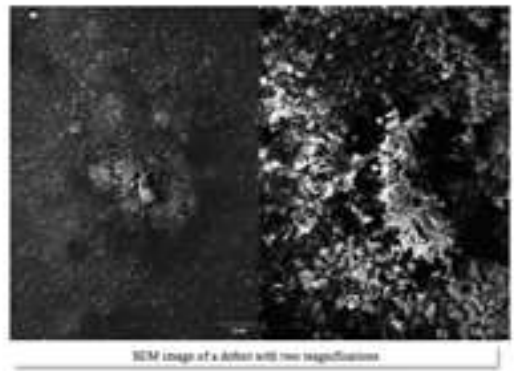}\\
  \caption{SEM image of a defect in a RPC chamber irradiated at the GIF under two magnifications. The defect is identified as NaF.}
  \end{center}
\end{figure}

\begin{figure}
\begin{center}
  \includegraphics[]{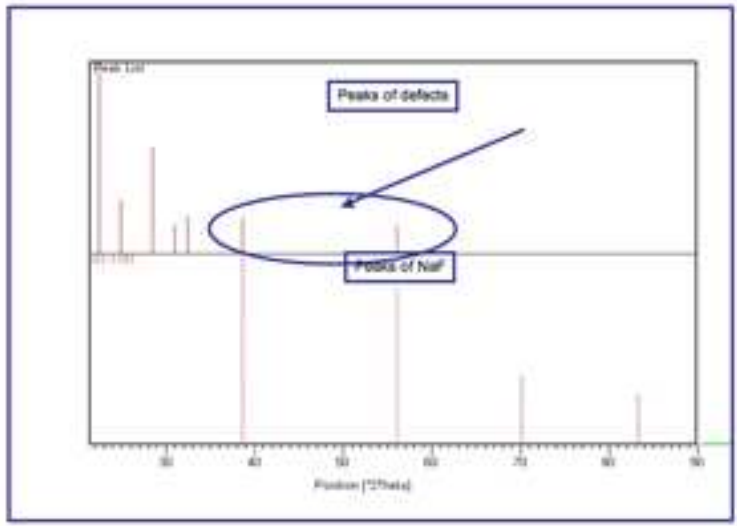}\\
  \caption{XRD spectrum of defects in a chamber irradiated at GIF. A very good match is observed for two characteristic peaks of  NaF.}
  \end{center}
\end{figure}

\begin{figure}
\begin{center}
  \includegraphics[]{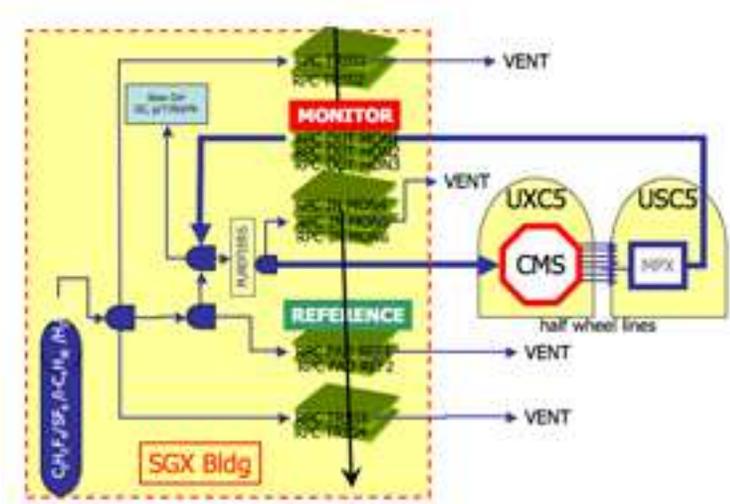}\\
  \caption{Gas gain monitoring system conceptual layout.}
  \end{center}
\end{figure}

\begin{figure}
\begin{center}
  \includegraphics[]{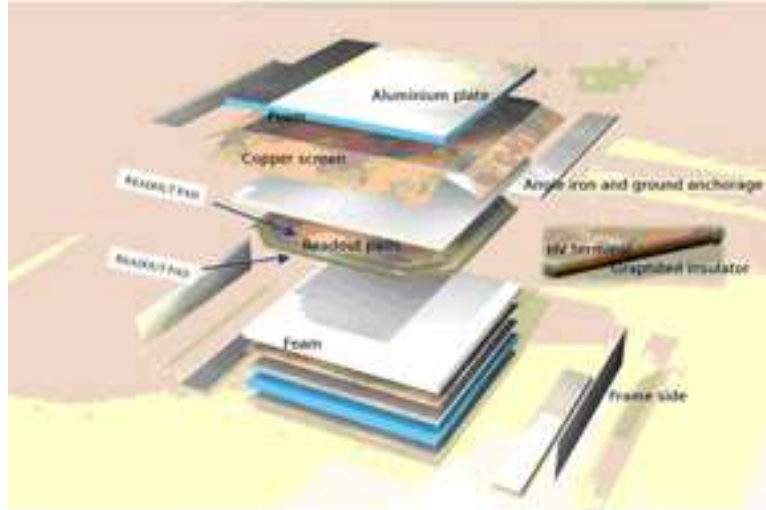}\\
  \caption{Exploded view of a 50cm x 50cm single gap which composes a subsystem of the gas gain monitoring system. Each single gap is read by two 45cm x 45cm pads on both sides, each one picking up negative and positive pulses from the avalanche developing in the gas gap upon the crossing of a cosmic ray track. }
  \end{center}
\end{figure}

\begin{figure}
\begin{center}
  \includegraphics[]{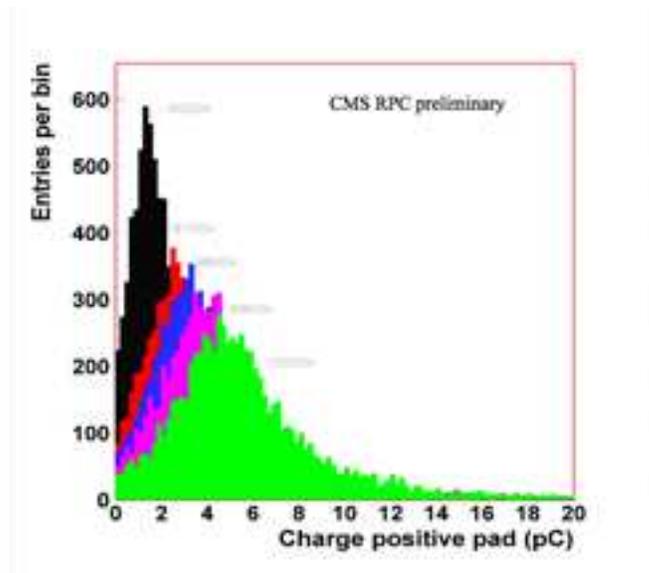}\\
  \caption{Charge distribution of single gap operated in avalanche regime and read by two 45cm x 45cm pads. The high voltage is increased from 9.5kV to 10kV. Pedestal is subtracted. }
  \end{center}
\end{figure}

\begin{figure}
\begin{center}
  \includegraphics[]{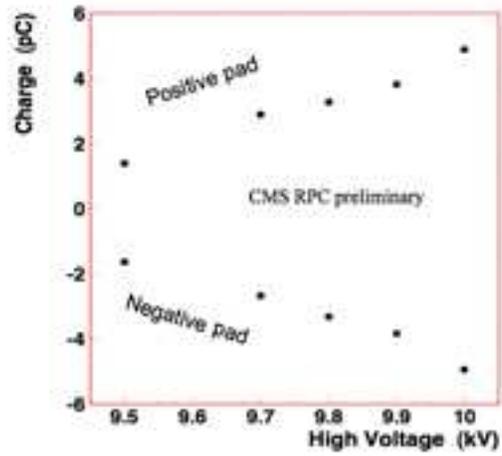}\\
  \caption{Dependence of peak values of charge distributions from high voltage}
  \end{center}
\end{figure}

\begin{figure}
\begin{center}
  \includegraphics[]{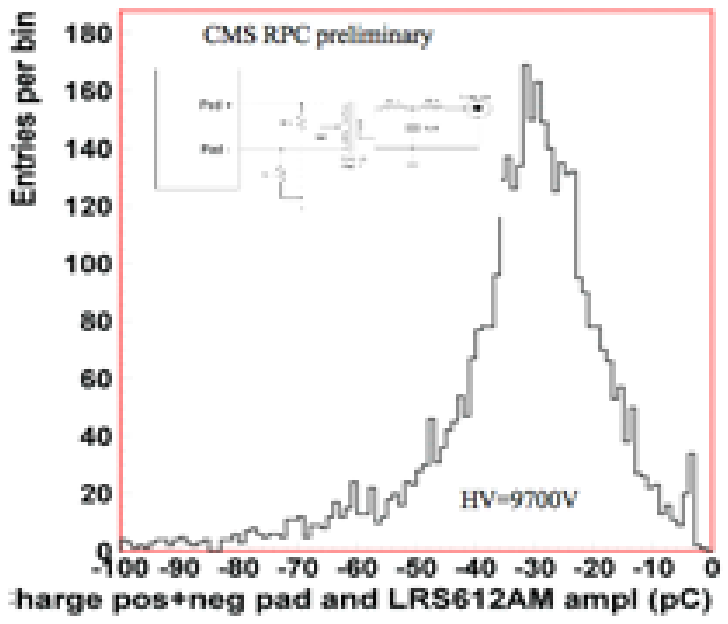}\\
  \caption{Charge distribution from the double pad readout scheme. A transformer (inset) sums positive and negative pads outputs. Transformer output is filtered and sent to NIM LeCroy amplifier. Pedestal is not subtracted and shown in arbitrary vertical scale at about -4pC.}
  \end{center}
\end{figure}

\begin{thebibliography}{0}

\bibitem
[1]	CMS Collaboration home page is /cms.cern.ch/.
\bibitem
[2]	G. Pugliese for the CMS RPC Collaboration, N24-3, this Conference.
\bibitem
[3]	M. Abbrescia et al., Nucl. Instr. And Meth. A533 (2994) 102-106; M.Abbrescia et al., Nucl. Instr. And Meth. A515 (2993) 342-347.
\bibitem
[4]	F.Hahn, CERN Closed Loop Gas System: Operating Instructions (unpublished).
\bibitem
[5]	M.Abbrescia et al., HF production in CMS Resistive Plate Chambers, submitted to NIM, presented by R.Guida at the RPC Conference, Seoul (Korea) 2005.
\bibitem
[6]	M.Abbrescia et al.,  Proposal for a systematic study of the CERN closed loop gas system used by the RPC muon detectors in CMS, Frascati preprint  LNF-06/26(IR).
\bibitem
[7]	C.Pucci, Master's Degree Thesis, University of Rome La Sapienza 2006 (in Italian). Frascati preprint LNF-06/31(Thesis).
\end{thebibliography}
\end{document}